\documentclass[aps,pra,reprint,onecolumn,notitlepage,eqsecnum,floatfix,nofootinbib,showkeys]{revtex4-2}

\usepackage{amsmath}
\usepackage{color}
\usepackage{graphicx}
\usepackage{hyperref}
\usepackage{multirow}
\usepackage{epigraph}

\newcommand{\bq}{\begin{eqnarray}}
\newcommand{\eq}{\end{eqnarray}}
\newcommand{\bqn}{\begin{eqnarray*}}
\newcommand{\eqn}{\end{eqnarray*}}
\newcommand{\bqs}{\begin{subequations}}
\newcommand{\eqs}{\end{subequations}}
\newcommand{\bw}{\begin{widetext}}
\newcommand{\ew}{\end{widetext}}
\newcommand{\xx}{{\bf x}}

\begin{document}
\title{Statistical Gravity through Affine Quantization}

\author{Riccardo Fantoni}
\email{riccardo.fantoni@scuola.istruzione.it}
\affiliation{Universit\`a di Trieste, Dipartimento di Fisica, strada
  Costiera 11, 34151 Grignano (Trieste), Italy}

\date{\today}

\begin{abstract}
I propose a possible way to introduce the effect of temperature (defined through the virial theorem) 
into Einstein's theory of general relativity. This requires the computation of a path integral on a 
10-dimensional {\sl flat} space in a four dimensional spacetime {\sl lattice}. Standard path 
integral Monte Carlo methods can be used to compute it.
\end{abstract}

\keywords{General Relativity; Einstein-Hilbert action; Statistical Physics; Path Integral; Monte Carlo; Virial Theorem}

\maketitle
\section{Introduction}
\label{sec:intro}

A still unsolved problem in physics is the formulation of a well defined theory unifying gravity and 
quantum mechanics. In this paper we propose a possible way to estimate thermal statistical effects 
on the fabric of Einstein's spacetime. Once an action for the Einstein field equations of general 
relativity is found we use it to construct a path integral formulation for statistical gravity which
will describe quantum effects at low temperatures. Recent progress on successful affine quantization 
\cite{Fantoni23b} of euclidean relativistic scalar field theories can become important in making 
this path integral mathematically well defined and accessible numerically, for example through the 
usual path integral Monte Carlo method \citep{Ceperley1995}. 

Some recent developments on gravity-quantum physics(-statistical physics) interplay are for example:

On 15 August 2016 Jeff Steinhauer, an experimental physicist at the Technion-Israel Institute of 
Technology in Haifa, has created an artificial black hole that seems to emit `Hawking radiation' on 
its own, from quantum fluctuations that emerge from its experimental set-up \cite{Castelvecchi2016}.
In a Bose-Einstein condensate (BEC) of {\bf rubidium cold atoms} he created an event horizon by 
accelerating the atoms until some were traveling at more than  $1 {\rm mm}/{\rm s}$ -- a supersonic 
speed for the condensate. On one side of his acoustical event horizon, where the atoms move at 
supersonic speeds, phonons became trapped. And when Steinhauer took pictures of the BEC, he found 
correlations between the densities of atoms that were an equal distance from the event horizon but 
on opposite sides. This demonstrates that pairs of phonons were entangled -- a sign that they 
originated spontaneously from the same quantum fluctuation and that the BEC was producing Hawking 
radiation.

Although quantum objects, like {\bf the proton}, are not conventionally considered within a unified 
physics framework that includes gravity, it is significant to note that it has recently been 
discovered that these particles are more dense than massively compact objects like neutron stars. 
Scientists at Jefferson Lab have measured pressures of 100 billion trillion trillion pascals at the 
proton's core -- about ten times greater than the pressure inside neutron stars. These are the same 
pressure gradients computed by Haramein et al. \cite{Haramein2023}, from first-principle 
considerations alone, on quantum vacuum fluctuations and gravitational effects at the proton scale, 
suggesting spacetime curvature alone is the source of mass and force.
The measurements by the Jefferson Lab team were achieved through innovative experiments that use 
pairs of photons to simulate gravitational interactions, allowing researchers to map the internal 
forces and pressures of the proton for the first time. The team found two distinct regions of 
specific pressures, like two phases within the proton, which was nearly exactly described in the 
study ``The Origin of Mass and the Nature of Gravity'' \cite{Haramein2023} by Haramein and the 
research 
team at the International Space Federation. A perfect example of how empirical research can reveal 
when a theory is on the right track, like when the Muonic measurements of the proton radius 
confirmed Haramein's prediction of a smaller proton radius than what was expected by the Standard 
Model, and now experimental data is revealing that, indeed, the mass-energy and pressures within the 
proton are sufficient that spacetime curvature must be a major consideration in the mass and binding 
forces that are observed for nucleons.

In 1974, Stephen Hawking predicted that quantum effects in the proximity of a black hole lead to the
emission of particles and black hole evaporation. At the very heart of this process lies a 
logarithmic phase singularity which leads to the Bose-Einstein statistics of Hawking radiation. An 
identical singularity appears in the elementary quantum system of the inverted harmonic oscillator. 
In Ref. \cite{Rozenman2024} the authors report the observation of the onset of this logarithmic 
phase singularity emerging at a horizon in phase space and giving rise to a Fermi-Dirac 
distribution. For this purpose, they utilize {\bf solitonic surface gravity water waves} and freely 
propagate an 
appropriately tailored energy wave function of the inverted harmonic oscillator to reveal the phase 
space horizon and the intrinsic singularities. Due to the presence of an amplitude singularity in 
this system, the analogous quantities display a Fermi-Dirac rather than a Bose-Einstein 
distribution.

In this work we try to carry to the extreme consequences the belief that the {\sl Wick rotation}, 
that allows to go back and forth between quantum physics and statistical physics in a 
non-general relativity framework, will continue to do so also in a (quantum) general relativity 
framework. This belief is 
based only on the grounds of a mathematical extension but allows us to reach the amusing physical 
interpretation of a fabric of space-time (the main actor of the new theory) which has thermal 
fluctuations itself. We will in fact introduce an effective temperature (field) and we will show its 
relationship with the absolute temperature of thermodynamics in the non-quantum non-relativistic 
limit.
 
\section{Einstein's field equations from a variational principle}
\epigraph{Sempre caro mi fu quest'ermo colle,\\
e questa siepe, che da tanta parte\\
dell'ultimo orizzonte il guardo esclude.}{\textit{Giacomo Leopardi\\ L' Infinito}}

The Einstein-Hilbert action in general relativity is the action that yields the Einstein field 
equations through the stationary-action principle. With the $(-\,+\,+\,+)$ metric signature, the 
gravitational part of the action is given as \cite{Hilbert-qg,Feynman-qg}
\bq
S=\frac{1}{2\kappa}\int R\sqrt{-g}\,\mathrm{d}^{4}x,
\eq
where $g=\det(g_{\mu \nu })$ is the determinant of the metric tensor matrix, $\sqrt{-g}$ is the 
scalar density, $\sqrt{-g}\,\mathrm{d}^{4}x$ is the invariant volume element,
$R$ is the Ricci scalar, and $\kappa =8\pi Gc^{-4}$ is the Einstein gravitational constant 
($G$ is the gravitational constant and $c$ is the speed of light in vacuum). If it converges, the 
integral is taken over the whole spacetime. If it does not converge, $S$
is no longer well-defined, but a modified definition where one integrates over arbitrarily large, 
relatively compact domains, still yields the Einstein equation as the Euler-Lagrange equation of the 
Einstein-Hilbert action. The action was proposed \cite{Hilbert-qg} by David Hilbert in 1915 as part 
of his application of the variational principle to a combination of gravity and electromagnetism.

The stationary-action principle then tells us that to recover a physical law, we must demand that 
the variation of this action with respect to the inverse metric be zero, yielding
\bq
0=\delta S&=&\int \left[{\frac {1}{2\kappa }}{\frac {\delta \left({\sqrt {-g}}R\right)}{\delta g^{\mu \nu }}}\right]\delta g^{\mu \nu }\,\mathrm {d} ^{4}x\\
&=&\int \left[{\frac {1}{2\kappa }}\left({\frac {\delta R}{\delta g^{\mu \nu }}}+{\frac {R}{\sqrt {-g}}}{\frac {\delta {\sqrt {-g}}}{\delta g^{\mu \nu }}}\right)\right]\delta g^{\mu \nu }{\sqrt {-g}}\,\mathrm {d} ^{4}x.
\eq
Since this equation should hold for any variation $\delta g^{\mu \nu }$, it implies that
\bq \label{eq:em}
{\frac {\delta R}{\delta g^{\mu \nu }}}+{\frac {R}{\sqrt {-g}}}{\frac {\delta {\sqrt {-g}}}{\delta g^{\mu \nu }}}=0
\eq
is the equation of motion for the metric field.

The variation of the Ricci scalar in Eq. (\ref{eq:em}) follows from varying the Riemann curvature tensor, and then the Ricci curvature tensor. The first step is captured by the Palatini identity
\bq
\delta R_{\sigma \nu }\equiv \delta {R^{\rho }}_{\sigma \rho \nu }=\left(\delta \Gamma _{\nu \sigma }^{\rho }\right)_{;\rho}-\left(\delta \Gamma _{\rho \sigma }^{\rho }\right)_{;\nu}.
\eq
Using the product rule, the variation of the Ricci scalar $R=g^{\sigma \nu }R_{\sigma \nu }$ then 
becomes,
\bq
{\begin{aligned}\delta R&=R_{\sigma \nu }\delta g^{\sigma \nu }+g^{\sigma \nu }\delta R_{\sigma \nu }\\&=R_{\sigma \nu }\delta g^{\sigma \nu }+\left(g^{\sigma \nu }\delta \Gamma _{\nu \sigma }^{\rho }-g^{\sigma \rho }\delta \Gamma _{\mu \sigma }^{\mu }\right)_{;\rho},\end{aligned}}
\eq
where we also used the metric compatibility $g^{\mu \nu }_{;\sigma}=0$, and renamed the 
summation indices $(\rho ,\nu )\rightarrow (\mu ,\rho )$ in the last term.
When multiplied by ${\sqrt {-g}}$, the term 
$\left(g^{\sigma \nu }\delta \Gamma _{\nu \sigma }^{\rho }-g^{\sigma \rho }\delta \Gamma _{\mu \sigma }^{\mu }\right)_{;\rho}$ becomes a total 
derivative, since for any vector $A^{\lambda }$ and any tensor density ${\sqrt {-g}}\,A^{\lambda }$, 
we have
\bq
{\sqrt {-g}}\,A_{;\lambda }^{\lambda }=\left({\sqrt {-g}}\,A^{\lambda }\right)_{;\lambda }=\left({\sqrt {-g}}\,A^{\lambda }\right)_{,\lambda }.
\eq
By Stokes' theorem, this only yields a boundary term when integrated. The boundary term is in 
general non-zero, because the integrand depends not only on $\delta g^{\mu \nu }$, but also on its 
partial derivatives 
$\partial _{\lambda }\,\delta g^{\mu \nu }\equiv \delta \,\partial _{\lambda }g^{\mu \nu }$. 
However, when the 
variation of the metric $\delta g^{\mu \nu }$ vanishes in a neighbourhood of the boundary or when 
there is no boundary, this term does not contribute to the variation of the action. Thus, we can 
forget about this term and simply obtain 
\bq \label{eq:vR}
{\frac {\delta R}{\delta g^{\mu \nu }}}=R_{\mu \nu }.		
\eq
at events not in the closure of the boundary.

The variation of the determinant in Eq. (\ref{eq:em}) requires Jacobi's formula, the rule for differentiating a determinant:
\bq
\delta g=gg^{\mu \nu }\delta g_{\mu \nu }.
\eq
Using this we get
\bq
\delta {\sqrt {-g}}=-{\frac {1}{2{\sqrt {-g}}}}\delta g={\frac {1}{2}}{\sqrt {-g}}\left(g^{\mu \nu }\delta g_{\mu \nu }\right)=-{\frac {1}{2}}{\sqrt {-g}}\left(g_{\mu \nu }\delta g^{\mu \nu }\right)
\eq
In the last equality we used the fact that from the symmetry of the metric tensor and 
$g_{\mu\nu}g^{\nu\mu}=\delta_\mu^\mu=4$ follows
\bq
g_{\mu \nu }\delta g^{\mu \nu }=-g^{\mu \nu }\delta g_{\mu \nu }
\eq
Thus we conclude that
\bq \label{eq:vg}
{\frac {1}{\sqrt {-g}}}{\frac {\delta {\sqrt {-g}}}{\delta g^{\mu \nu }}}=-{\frac {1}{2}}g_{\mu \nu }.
\eq

Now that we have all the necessary variations at our disposal, we can insert Eq. (\ref{eq:vg}) and 
Eq. (\ref{eq:vR}) into the equation of motion (\ref{eq:em}) for the metric field to obtain
\bq
G_{\mu\nu}\equiv R_{\mu \nu }-{\frac {1}{2}}g_{\mu \nu }R=0,
\eq
which is the Einstein field equations in vacuum.

Moreover, since Einstein's tensor $G_{\mu\nu}$ appears from a variational 
principle:
\bq
\frac{\delta S}{\delta g^{\mu\nu}}=\frac{1}{2\kappa}\sqrt{-g}\left(R_{\mu\nu}-
\frac{1}{2}g_{\mu\nu}R\right)=\frac{1}{2\kappa}G_{\mu\nu},
\eq
its covariant divergence is necessarily zero \cite{Feynman-qg}.

Matter or electromagnetic fields will produce a curvature of spacetime. In order to take this into 
account it is necessary to add a term ${\cal L}_F$ as follows,
\bq \label{eq:EHA}
S=\int \left(\frac{1}{2\kappa}R+{\cal L}_{\mathrm {F}}\right)\sqrt{-g}\,\mathrm{d}^{4}x.
\eq
The equations of motion coming from the stationary-action principle now become
\bq \label{eq:efe}
G_{\mu\nu}\equiv R_{\mu \nu }-{\frac {1}{2}}g_{\mu \nu }R=\kappa T_{\mu\nu},
\eq
where 
\bq \label{eq:set}
T_{\mu\nu}={\frac {-2}{\sqrt {-g}}}{\frac {\delta ({\sqrt {-g}}{\mathcal {L}}_{\mathrm {F} })}{\delta g^{\mu \nu }}}=-2{\frac {\delta {\mathcal {L}}_{\mathrm {F} }}{\delta g^{\mu \nu }}}+g_{\mu \nu }{\mathcal {L}}_{\mathrm {F} },
\eq
is the stress-energy tensor and $\kappa =8\pi G/c^{4}$ has been chosen such that the 
non-relativistic limit yields the usual form of Newton's gravity law.

\section{Path integral formulation of statistical gravity}
\epigraph{[$\ldots$] e il suon di lei. Così tra questa\\
immensit\`a s'annega il pensier mio:\\
e il naufragar m'\`e dolce in questo mare.}{\textit{Giacomo Leopardi\\ L' Infinito}}

Then the action for Einstein's theory of general relativity is one for a particular field theory 
where the field is the metric tensor $g_{\mu\nu}(x)$ a symmetric tensor with 10 independent 
components, each of which is a smooth function of 4 variables. We will indicate all this components 
with the notation $\{g\}(x)$. We will also work in euclidean time $x^0\equiv ct\to ict$ so that the 
metric signature becomes $(+\,+\,+\,+)$.

The thermal average of an observable ${\cal O}[\{g\}(x)]$ will then be given 
by the following expression
\bq \label{eq:EV}
\langle{\cal O}\rangle=\frac{\int{\cal O}[\{g\}(x)]\exp(-\upsilon S)\;{\cal D}^{10}\{g\}(x)}{\int\exp(-\upsilon S)\;{\cal D}^{10}\{g\}(x)},
\eq
so that $\langle 1\rangle=1$. Here $1/\upsilon$ is a positive constant of dimension of energy times 
length, $ct\in [0,\beta[$ where $\beta=1/\tilde{k}_B\tilde{T}$, $\tilde{k}_B$ is a Boltzmann 
constant of dimensions of one divided by length and by degree Kelvin, and $\tilde{T}$ an {\sl 
effective} temperature in degree Kelvin. Since the thermal average involves taking a trace we must 
have $g_{\mu\nu}(t+\beta/c,\xx)=g_{\mu\nu}(t,\xx)$, where we denote with $\xx\equiv (x^1,x^2,x^3)$ a 
spatial point. $S$ is the Einstein-Hilbert action of Eq. (\ref{eq:EHA}). We will also require 
periodic spatial boundary conditions on the finite volume 
$\Omega\subset {\mathrm I}\!{\mathrm R}^3$ which is the closest thing to a formal thermodynamic 
limit. As usual we will use 
${\cal D}^{10}\{g\}(x)\equiv\prod_{x}d^{10}\{g\}(x)$ and the functional 
integrals will be calculated on a lattice using the path integral Monte Carlo (PIMC) method
\cite{Ceperley1995}. Moreover we will choose $d^{10}\{g\}(x)\equiv\prod_{\mu\le\nu}dg^{\mu\nu}(x)$
where the 10-dimensional space of the 10 independent components of the symmetric metric tensor is 
assumed to be flat.

The real metric fields $\{g\}$ take the values $\{g\}(x)$ on each site of a periodic 
$4$-dimensional lattice of lattice spacing $a$, the ultraviolet cutoff, spatial periodicity 
$L=Na$, and temporal periodicity $\beta=N_0a$. The metric path is a closed loop on a $4$-dimensional 
surface of a $5$-dimensional $\beta$-cylinder.
We are interested in reaching the continuum limit by taking $L=Na$ fixed and letting 
$N\to\infty$ at fixed volume $L^3$. As for the temporal periodicity we can 
initially treat it on the same footing of the spatial periodicity taking $\beta=L$ and $N_0=N$. 
On the other hand we can distinguish the two periodicities allowing the effective 
temperature $\tilde{T}=1/\tilde{k}_B\beta$ to vary so that the number of discretization 
points for the imaginary time interval $[0,\beta[$ will be $N_0=\beta/a$.  
The PIMC simulation (the computer experiment) may use the Metropolis algorithm 
\cite{Kalos-Whitlock,Metropolis} to calculate the discretized version of Eq. (\ref{eq:EV}) 
which is a $10N^3N_0$ multidimensional integral. So clearly the computational resources 
needed to perform a given simulation depends critically on $a$. We may use natural Planck units 
$c = \hbar = k_B = 1$. The simulation is started from any initial condition. 
One MC step consists in a random displacement of each one of the $10N^3N_0$ 
independent components of $\{g\}$ as follows
\bq
g^{\mu\nu}\rightarrow g^{\mu\nu}+(2\eta-1)\delta,
\eq
where $\eta$ is a uniform pseudo random number in $[0,1]$ and $\delta$ is the amplitude of the 
displacement. Each one of these $10N^3N_0$ moves is accepted if 
$\exp(-\upsilon\Delta S)>\eta$, where $\Delta S$ is the change in the 
action due to the move, and rejected otherwise. The amplitude $\delta$ is chosen in such a way 
to have acceptance ratios as close as possible to $1/2$ and is kept constant during the 
evolution of the simulation. One simulation consists of a 
large number of $p$ steps. The statistical error on the average $\langle{\cal O}\rangle$ will then 
depend on the correlation time, $\tau_{\cal O}$, necessary to decorrelate 
the property ${\cal O}$ and will be determined as 
$\sqrt{\sigma_{\cal O}^2\tau_{\cal O}/[p10N^3N_0]}$, where 
$\sigma_{\cal O}^2$ is the intrinsic variance for ${\cal O}$.
Our estimate of the path integrals will be generally subject 
to three sources of numerical  uncertainties: The one due to the {\sl statistical error}, 
the one due to the {\sl spacetime discretization}, and the one due to the 
{\sl finite-size effects}. Of these, the statistical
error scales like $P^{-1/2}$ where $P=p10N^3N_0$ is the computer time, the discretization of 
spacetime is responsible for the distance from the continuum limit (which corresponds to a 
lattice spacing $a\to 0$), and the finite-size effects stems from the necessity to approximate 
the infinite spacetime system with one in a hypertorus of volume $L^3\beta$.

This makes sense since the Einstein theory does not predict a curvature of spacetime due to 
temperature so it does not take care of the virial theorem of statistical physics. According to the 
virial theorem the temperature of a portion of spacetime should be related to kinetic energy which 
in turn should enter the stress energy tensor and be responsible for the curvature of spacetime 
according to Eq. (\ref{eq:efe}). Here we are thinking at ${\cal L}_F$ as a `potential energy' 
density of interaction of the metric field. Therefore we should take care of the effect 
of temperature in some alternative way. Here we propose to use the usual formalism of statistical 
physics to estimate thermal averages $\langle\ldots\rangle$ of observables that depend on the metric 
tensor like for example the Riemann curvature tensor or the Christoffel symbols. This requires to 
use a path integral over a 10-dimensional space like in Eq. (\ref{eq:EV}). Then, both the curvature 
of spacetime and the geodesic equation of motion of a point particle will be influenced by 
temperature and one should replace them with their thermally averaged versions more correctly. In 
the classical high temperature limit applying a time average to the contracted Einstein field 
equations (\ref{eq:efe}), $-R=\kappa T_\mu^\mu$, assuming ${\cal L}_F$ independent of $g^{\mu\nu}$ 
so that $T_\mu^\mu=4{\cal L}_F$, and replacing the time average with the ensemble 
average of Eq. (\ref{eq:EV}) in the first member, we soon reach (see the Appendix) the following 
relation $-2\kappa/\beta\bar{\upsilon}=\frac{\kappa}{2}\langle T_\mu^\mu\rangle_t$, with 
$\langle\ldots\rangle_t=\lim_{\tau\to\infty}\int_0^\tau\ldots dt/\tau$ the time average, 
$\upsilon'$ another constant of dimension length squared divided by energy, $\kappa$ has 
dimensions of length divided by energy, and the stress energy tensor elements have dimensions 
of an energy density. This gives a temperature definition 
$-\tilde{k}_B\tilde{T}=\frac{\bar{\upsilon}}{4}\langle T_\mu^\mu\rangle_t$.

In Eq. (\ref{eq:EV}) we assumed a constant temperature throughout the whole accessible spacetime. 
This can be a too restrictive condition and it could be necessary to think about a temperature 
scalar field $\tilde{T}(x)$ which gives the value of the temperature in a neighborhood of a given 
event $x$. Actually we are bound to choose $\tilde{T}(\xx)$ as the temperature in a neighborhood of 
a given spatial point $\xx$ since we must require $x^0\in [0,\beta(\xx)[$. And the temperature field 
$\tilde{T}(\xx)=1/\tilde{k}_B\beta(\xx)$ could be available experimentally.

In Eq. (\ref{eq:EV}) we assumed that what gives rise to the thermal average is the product of the 
10 independent components of the tensor path integral 
${\cal I}^{\mu\nu}=\int \ldots {\cal D}g^{\mu\nu}(x)$. This may be regarded as a rather arbitrary 
receipt but it gives us something that is numerically amenable and directly accessible through path 
integral Monte Carlo. Moreover it gives the correct classical high temperature limit 
$\langle{\cal O}\rangle\to\langle{\cal O}_c\rangle$ where ${\cal O}_c$ is the value of the 
observable at the metric of the solution of the classical Einstein field equations (\ref{eq:efe}).

In order to make precise these arguments and to treat correctly the integral in Eq. (\ref{eq:EV})  
we immediately recognize that it is necessary a splitting of the time component from the 3 spatial 
components in the spirit of the Arnowitt–Deser–Misner (ADM) foliation formalism. In this respect 
affine quantization as explained in Ref. \cite{Fantoni23b} may become useful.

\section{Conclusions}

We propose a way to include into Einstein's general relativity theory the effect of temperature. 
This involves the use of a path integral on a 10-dimensional flat space in a four dimensional 
spacetime lattice hypertorus. It may prove necessary introduce a temperature scalar field function 
of the position within the spatial volume under exam. From the definition of the thermal average we
will have that at large temperature, i.e. small $\beta$, the quantum effects of statistics on 
the spacetime fabric will be irrelevant. On the other hand the quantum effects will become important 
at low temperatures, i.e. large $\beta$. This means that the quantum effects are negligible whenever
the path in the metric components extends over a short imaginary time interval, i.e. the hypertorus 
`radius' around the time curled dimension is small and it reduces to something like a round 
closed hyper`string' whose thickness varies along its length. We will call our theory the FEBB from 
the initials of the surnames of the protagonists of the 3 main advances in theoretical physics: 
Albert Einstein for physics of gravitation, Ludwig Boltzmann for statistical physics, and Niels Bohr 
for quantum physics.

The internal consistency of our FEBB theory requires a definition of `temperature' as 
\bq
\tilde{T}=-\frac{\bar{\upsilon}}{4\tilde{k}_B}\langle T_\mu^\mu\rangle_t,
\eq
where $\tilde{k}_B$ is a Boltzmann constant with dimensions of one divided by length and by degree 
Kelvin.

Affine quantization as explained in Ref. \cite{Fantoni23b} may become useful to make these arguments 
precise and the physical meaning of the factor $\bar{\upsilon}$.

We conclude observing that for an ideal fluid in thermodynamic equilibrium in imaginary time
$T_\mu^\mu=5P+\rho c^2$ where $P$ is the pressure and $\rho$ the mass-energy density, so that in the 
non-relativistic and non-quantum limits we must require that 
\bq \label{eq:eos}
-4\frac{\tilde{k}_B\tilde{T}}{\bar{\upsilon}}-\rho c^2\to 5 n k_BT,
\eq
where $k_B$ and $T$ are the usual Boltzmann constant and absolute temperature respectively. So to
recover the usual ideal gas equation of state $P=nk_BT$ where 
$P=\langle P\rangle_t=\langle P\rangle$ is pressure and 
$n=\langle n\rangle_t=\langle n\rangle$ is number density. Note that in Eq. (\ref{eq:eos}) 
$\bar{\upsilon}>0$, $\rho c^2>0$, and the right hand side is also a positive quantity. 
But thanks to the circular boundary conditions around the imaginary time we are free to choose 
$\beta<0$. And this should be the case (see footnote \ref{fn}) since from the trace of Einstein 
field equations follows, in imaginary time, $-R=\kappa T^\mu_\mu>0$, i.e. $R<0$. Since in the 
non-relativistic limit, $c\to\infty$,  we have in 
imaginary time $\rho c^2\to nm_0c^2-n\frac{1}{2}m_0v^2$ with $m_0$ the rest mass of the particles 
and $v$ their velocities. We will find for the equipartition theorem 
$\langle\rho c^2\rangle\to nm_0c^2-\frac{3}{2}k_BT$ we require, in the non-relativistic 
($c\to\infty$) non-quantum ($\tilde{T}\to\infty$) limits
\bq \label{eq:nq-nr}
4\frac{\tilde{k}_B\tilde{T}}{\tilde{\upsilon}}-nm_0c^2\to\frac{7}{2}nk_BT,
\eq
where we called $\tilde{\upsilon}=-\bar{\upsilon}$.
We then see that $\tilde{T}$ should be chosen proportional to the number density $n$. This is a 
necessary constraint if we require our initial belief that a Wick rotation will bring us from
Quantum Gravity to Statistical Gravity. Our Eq. (\ref{eq:nq-nr}) makes sense since 
infinity minus infinity is an indeterminate form and can be equal to a finite result.
 
We ask the reader, could our effective (obscure) temperature give some hints in the search for the 
{\bf dark matter} in the universe. Also, could our new theory explain the recent findings about the 
{\bf tension} in the universe \cite{Riess2024}. 
 
\acknowledgments
I thank the two women of my life, the first for giving me the chance of learning path integral and 
general relativity and the second for giving me the chance of learning affine quantization.

\appendix
\section{Derivation of the virial theorem}

Taking the thermal average of the trace of the Einstein field equations \ref{eq:efe} we find
\bq \label{eq:avtefe}
\langle -R\rangle=\kappa\langle T_\mu^\mu\rangle.
\eq
Now we notice that we can evaluate the left hand side as follows
\bq
\langle -R\rangle &=&\frac{\int(-R)\exp(-\upsilon S)\;{\cal D}^{10}\{g\}(x)}{\int\exp(-\upsilon S)\;{\cal D}^{10}\{g\}(x)}.
\eq
In the high temperature limit we will have, from Eq. (\ref{eq:EHA}),
\bq
\upsilon S\approx\beta\bar{\upsilon}\left(\frac{1}{2\kappa}R+{\cal L}_{\mathrm {F}}\right).
\eq
Note that this is only an approximation and we would more correctly need a 3+1 splitting of space 
from imaginary time. If we do not worry about these details, for the time being, we reach the 
following result
\bq
\langle -R\rangle &\approx&\frac{2\kappa}{\bar{\upsilon}}\frac{(d/d\beta)\int\exp(-\upsilon S)\;{\cal D}^{10}\{g\}(x)}{\int\exp(-\upsilon S)\;{\cal D}^{10}\{g\}(x)}+2\kappa\langle{\cal L}_F\rangle.
\eq
Now, assuming ${\cal L}_F$ independent from the metric tensor, we find from Eq. (\ref{eq:set}) that
$T_\mu^\mu=4{\cal L}_F$ and also
\bq
\langle -R\rangle &\approx&\frac{2\kappa}{\bar{\upsilon}}\frac{(d/d\beta)\int\exp(-\upsilon S)\;{\cal D}^{10}\{g\}(x)}{\int\exp(-\upsilon S)\;{\cal D}^{10}\{g\}(x)}+\frac{\kappa}{2}\langle T_\mu^\mu\rangle\\
&=&-\frac{2\kappa}{\beta\bar{\upsilon}}+\frac{\kappa}{2}\langle T_\mu^\mu\rangle,
\eq
where in the last equality we took profit of the fact that ${\cal L}_F$ is independent from the 
metric tensor components $\{g\}$ so that its contribution simplifies from the two path integrals in 
the numerator and in the denominator \footnote{Note that 
$\int_0^\infty e^{-\alpha y}\,dy=\frac{1}{\alpha}$ when $\alpha>0$. So that 
$[(d/d\alpha)\int_0^\infty e^{-\alpha y}\,dy]/\int_0^\infty e^{-\alpha y}\,dy=-\frac{1}{\alpha}$. 
Moreover in imaginary time we have a negative Ricci scalar curvature, i.e. $R<0$, so that we must 
have $\beta<0$. \label{fn}}.

Using this result into Eq. (\ref{eq:avtefe}) we reach the following result
\bq
-\frac{2\kappa}{\beta\bar{\upsilon}}+\frac{\kappa}{2}\langle T_\mu^\mu\rangle\approx\kappa\langle T_\mu^\mu\rangle,
\eq
or
\bq
-\frac{2\kappa}{\beta\bar{\upsilon}}\approx\frac{\kappa}{2}\langle T_\mu^\mu\rangle.
\eq
which gives the desired result upon replacing the thermal average with the time average 
introduced in the main text $\langle\ldots\rangle\to\langle\ldots\rangle_t$.

\section*{Author declarations}
\subsection*{Conflict of interest}
The author has no conflicts to disclose.

\section*{Data availability}
The data that support the findings of this study are available from the 
corresponding author upon reasonable request.
\bibliography{qg}

\end{document}